\documentclass[a4paper,10pt]{article}
\usepackage{adjustbox}
\usepackage{placeins}
\usepackage{siunitx}
\usepackage{authblk}
\usepackage{graphicx}
\usepackage{booktabs}
\usepackage{amsmath}
\usepackage{xcolor}
\usepackage[colorlinks]{hyperref}
\usepackage{url}
\usepackage[utf8]{inputenc}
\DeclareUnicodeCharacter{00B0}{\degree}

\def\tmp#1#2#3{%
  \definecolor{Hy#1color}{#2}{#3}%
  \hypersetup{#1color=Hy#1color}}
\tmp{link}{HTML}{800006}
\tmp{cite}{HTML}{2E7E2A}
\tmp{file}{HTML}{131877}
\tmp{url} {HTML}{8A0087}
\tmp{menu}{HTML}{727500}
\tmp{run} {HTML}{137776}
\def\tmp#1#2{%
  \colorlet{Hy#1bordercolor}{Hy#1color#2}%
  \hypersetup{#1bordercolor=Hy#1bordercolor}}
\tmp{link}{!60!white}
\tmp{cite}{!60!white}
\tmp{file}{!60!white}
\tmp{url} {!60!white}
\tmp{menu}{!60!white}
\tmp{run} {!60!white}

\begin{document}
\title{Multi-modal microscopy imaging with the OpenFlexure Delta Stage}

\author[1]{Samuel McDermott\thanks{\texttt{sjm263@cam.ac.uk}}}
\author[1]{Filip Ayazi}
\author[2]{Joel Collins}
\author[2]{Joe Knapper}
\author[2]{Julian Stirling}
\author[2]{Richard Bowman}
\author[1]{Pietro Cicuta}

\affil[1]{Department of Physics, University of Cambridge, UK}
\affil[2]{Department of Physics, University of Bath, UK}
\maketitle


\begin{abstract}
    Microscopes are vital pieces of equipment in much of biological research and medical diagnostics. However, access to a microscope can represent a bottleneck in research, especially in lower-income countries. `Smart' computer controlled motorized microscopes, which can perform automated routines or acquire images in a range of modalities are even more expensive and inaccessible. Developing low-cost, open-source, smart microscopes enables more researchers to conceive and execute optimized or more complex experiments.  Here we present the OpenFlexure Delta Stage, a 3D-printed microscope designed for researchers.  Powered by the OpenFlexure software stack, it is capable of performing automated experiments. The design files and assembly instructions are freely available under an open licence.  Its intuitive and modular design---along with detailed documentation---allows researchers to implement a variety of imaging modes with ease. The versatility of this microscope is demonstrated by imaging biological and non-biological samples (red blood cells with Plasmodium parasites and colloidal particles in brightfield, epi-fluorescence, darkfield, Rheinberg and differential phase contrast. We present the design strategy and choice of tools to develop devices accessible to researchers from lower-income countries, as well as the advantages of an open-source project in this context. This microscope, having been open-source since its conception, has already been built and tested by researchers around the world, promoting a community of expertise and an environment of reproducibility in science.
\end{abstract}

\section{Introduction}
\begin{figure}[!t]
    \centering
    \includegraphics[width=0.5\linewidth]{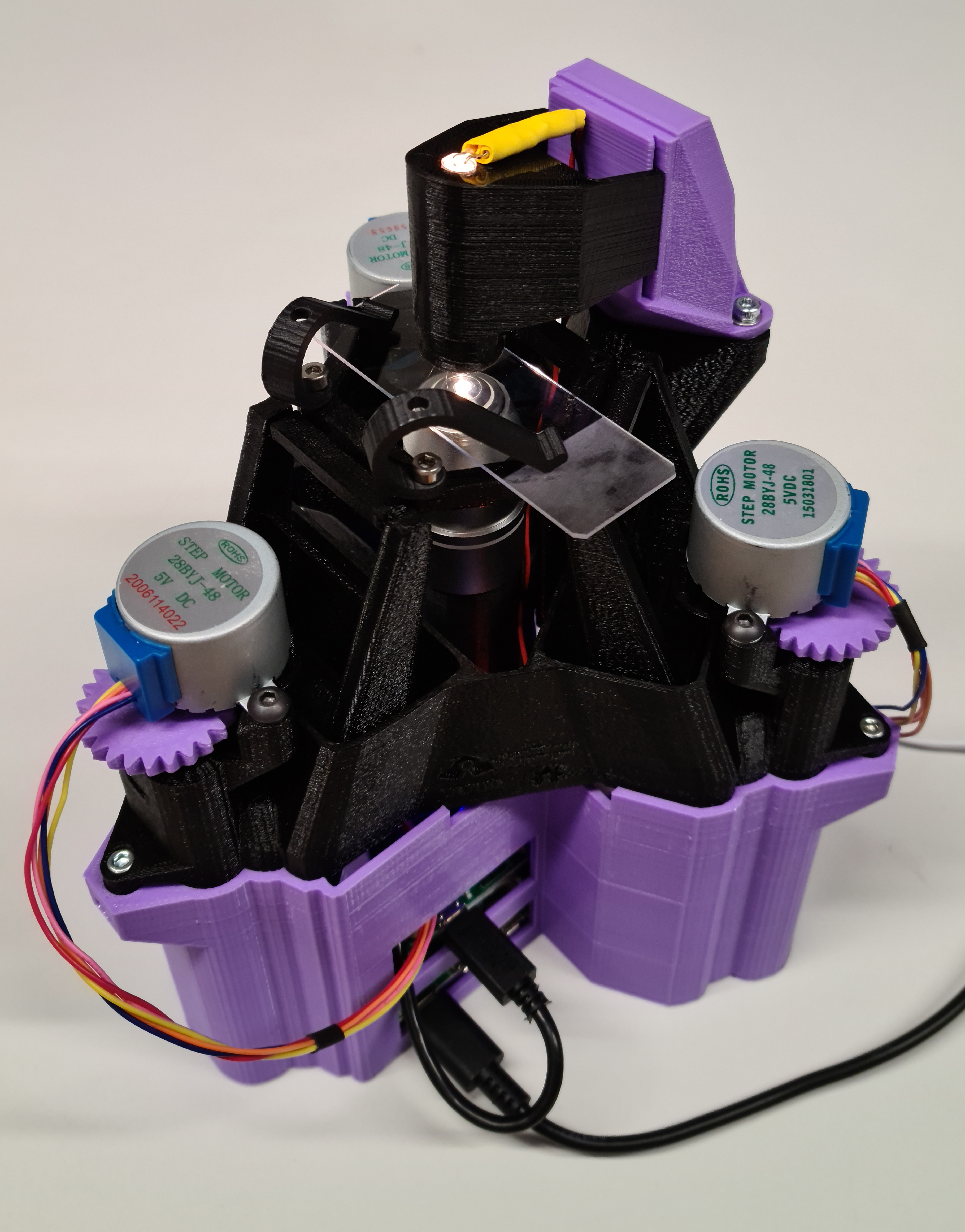}
    \caption{The OpenFlexure Delta Stage's compact and well-tested design is suitable for a laboratory bench or microbiological safety cabinet.  Here it is shown imaging a blood smear on a microscope slide. It is majority 3D printed, and powered using a Raspberry Pi. Its `delta' geometry stage allows the sample to be moved independently of the optics in $x$, $y$, and $z$.}
    \label{fig:delta-stage-photo}
\end{figure}

One of the most common scientific instruments,  particularly in biological research, is the optical microscope. Access to a microscope capable of multiple imaging modalities and automation is often a limiting factor even in well funded laboratories, where an advanced unit might be shared across many users. It is particularly acute in laboratories of lower income countries, limiting the scope and ambition of research projects.  The OpenFlexure Delta Stage (OFDS), shown in Figure~\ref{fig:delta-stage-photo}, aims to provide a low-cost microscope enabling a range of imaging modalities, in an open format that can be easily modified, and with parts that are mostly 3D printed enabling local manufacturing and repair.

The OFDS stems as a variant of the OpenFlexure Microscope (OFM), an open source project that has evolved over many years,  recently overviewed in~\cite{Collins2020}.  OFDS uses the same flexure properties provided by 3D printed plastics, but a different orientation of the actuators compared to OFM means that now the stage can move in 3 dimensions independently of the optics.  In contrast to the OpenFlexure Microscope's `Cartesian' geometry, where the stage moves in $x$ and $y$ and the optics move in $z$, the stage in the OFDS moves independently of the other optics modules, in each of $x,\,y$ and $z$. This makes it particularly suitable for heavier optics and more complex optical systems.  A early version of this 3D printed `delta robot' design~\cite{Rocholl2012} was previously used for optical projection tomography `OptiJ'~\cite{VallejoRamirez2019}, but this is the first time it has been developed into a self-contained microscope.

This paper demonstrates the versatility and simplicity of using the OFDS to implement  a wide range of imaging modes, as outlined in~Table~\ref{tab:imagingmodes}.  These microscopes are a self contained units, they can be built using common tools and parts, have comprehensive build instructions and are controlled using intuitive software. This makes  the unit  suitable both as an instrument in itself, for a variety of imaging applications, or for integration with other experimental instruments.  Low-cost, open-source, optical systems such as the UC2~\cite{Diederich2020}, Microscopi~\cite{Wincott2021}, and OctoPi~\cite{Li2019a} designs are becoming a popular way of increasing  accessibility to optical experiments, lowering the initial barrier for users to tailor their microscopy needs to their experiments. The open-source nature of the OFDS has already led to adaptors to link different modules and systems, such as using the OFDS with the UC2 system~\cite{Eisenstein2021}.

Throughout the design phase we have worked closely with biologists and physicists with a range of budgets and experience to create a microscope that would have significant functionalities whilst keeping in mind ease of use. The OFDS is compact enough to fit on a lab bench, such that it can become an everyday tool for  quickly checking a sample, digital capture of images, or can be built up as multiple units to enable a high throughput workflow.

\begin{table*}[b]
    \centering
    \caption{The OpenFlexure Delta Stage is capable of multiple imaging modes. The key imaging modes laid out in this table are described in this paper.}
    \label{tab:imagingmodes}
    \begin{adjustbox}{center}\resizebox{1.5\textwidth}{!}{\begin{tabular}{@{}llllll@{}}
                \toprule
                \textbf{}            & \textbf{Brightfield}                  & \textbf{Epi-fluorescence}             & \textbf{Darkfield}                     & \textbf{Rheinberg}                     & \textbf{Differential Phase Contrast}   \\ \midrule
                Description          & Section~\ref{sec:brightfield}         & Section~\ref{sec:epi-fluorescence}    & Section~\ref{sec:darkfield}            & Section~\ref{sec:rheinberg}            & Section~\ref{sec:dpc}                  \\
                Illumination module  & Transmission illumination             & Reflection illumination               & LED array                              & LED array                              & LED array                              \\
                Optics module        & Transmission optics module            & Reflection optics module              & Transmission optics module             & Transmission optics module             & Transmission optics module             \\
                Cross section figure & Figure~\ref{fig:all_ray_diagrams} (a) & Figure~\ref{fig:all_ray_diagrams} (b) & Figure~\ref{fig:all_ray_diagrams} (c)  & Figure~\ref{fig:all_ray_diagrams} (c)  & Figure~\ref{fig:all_ray_diagrams} (c)  \\
                Sample images        & Figure~\ref{fig:brightfield-results}  & Figure~\ref{fig:fluorescence-results} & Figure~\ref{fig:led_sample_images} (b) & Figure~\ref{fig:led_sample_images} (c) & Figure~\ref{fig:led_sample_images} (d) \\ \bottomrule
            \end{tabular}}
    \end{adjustbox}
\end{table*}

\begin{figure*}[t!]
    \centering
    \includegraphics[width=\linewidth]{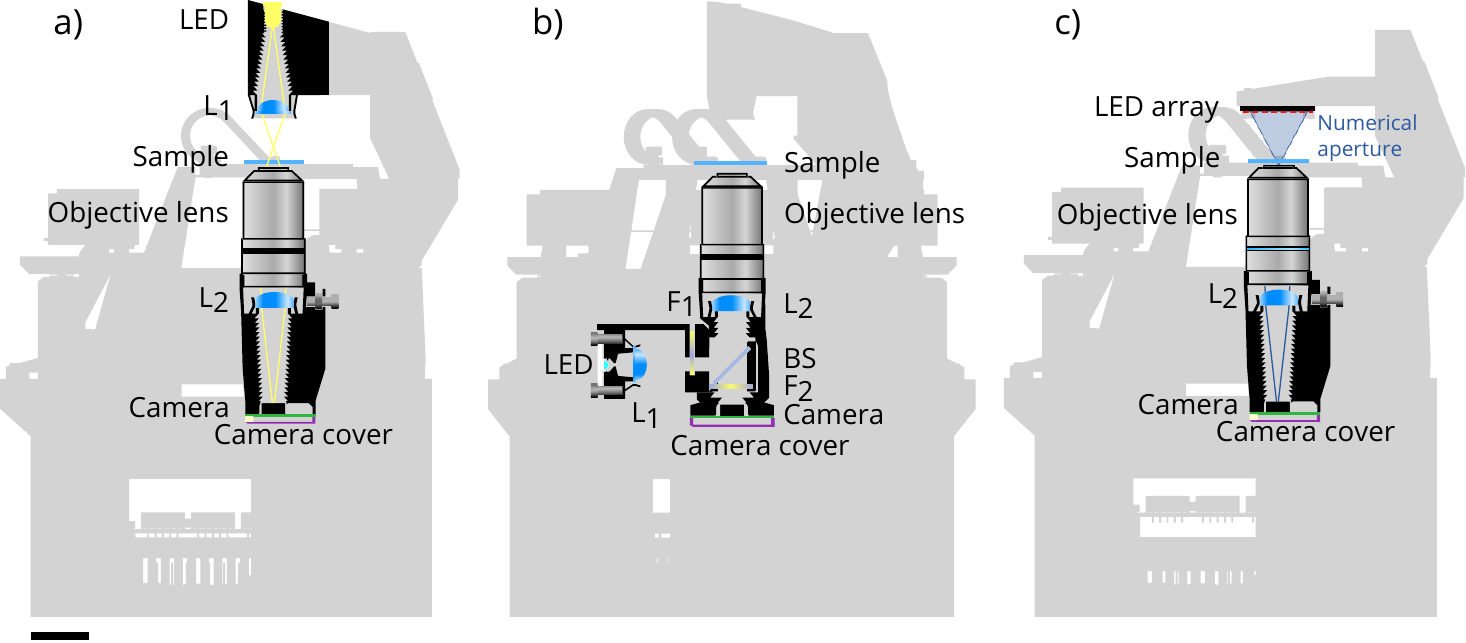}
    \caption{The OpenFlexure Delta Stage has three modular setups for performing different imaging modes. These cross-sections show the three main optics setups. (a) Brightfield (transmission) imaging. (b) Reflection imaging. (c) LED array imaging.  ( L\textsubscript{1}: Condenser lens, L\textsubscript{2}: `Tube length correction' lens, BS: Beamsplitter, F\textsubscript{1}: Emission filter, F\textsubscript{2}: Excitation filter) Scale bar 2cm.}
    \label{fig:all_ray_diagrams}
\end{figure*}

\section{Optics}
The optics of the OFDS  are designed to be modular.  There are three optical setups for imaging, illustrated with cross sectional diagrams in Figure~\ref{fig:all_ray_diagrams}: (a) Brightfield (transmission);  (b) Reflection; (c) Computational illumination with a LED array.

\begin{figure}[t!]
    \centering
    \includegraphics[width=0.8\linewidth]{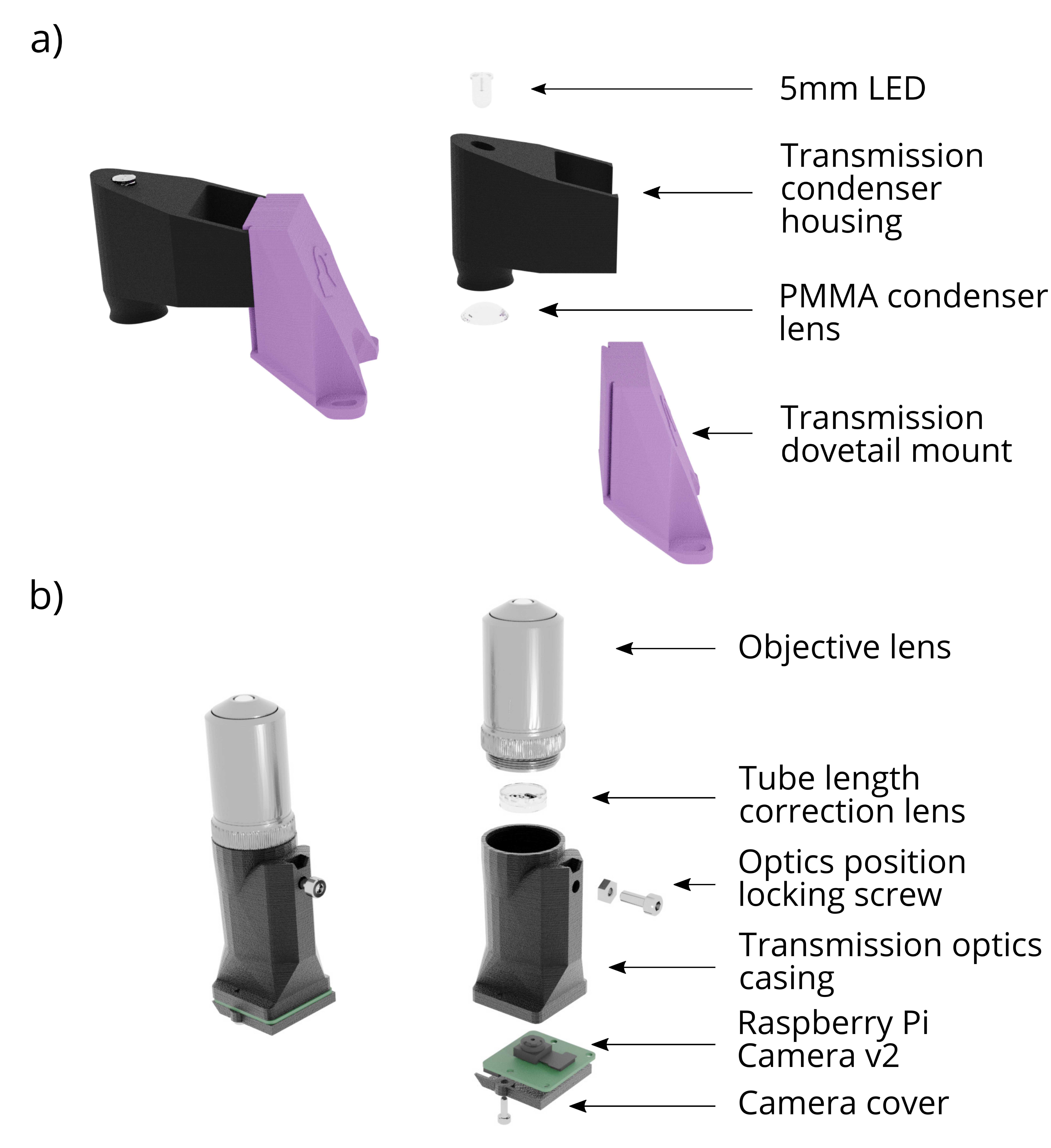}
    \caption{The transmission mode optics module is constructed from 3D printed parts and common optical components.   This computer generated render shows the assembly of the transmission (a) illumination and (b) optics modules.}
    \label{fig:transmission_illumination_assembly}
\end{figure}

\subsection{Brightfield (transmission) illumination}
\label{sec:brightfield}

\subsubsection{Optics module}

The optics module (Figure~\ref{fig:transmission_illumination_assembly}(a)) is a single printed part that holds an objective lens, a `tube length correction' lens and the imaging sensor in position.  It is attached to the main body with a wedge-shaped mount, allowing initial coarse adjustment of its vertical position.   The objective lens is a standard RMS threaded objective, which is screwed into the top of the optics module.  This type of lens provides high-quality imaging, suitable for scientific use, and is readily available from various manufacturers in a range of magnifications ($4\times-100\times$) and numerical apertures (and prices). It is recommended to use Plan or Semi-Plan corrected lenses up to $40\times$ and Plan corrected or better at $100\times$ magnification.

The optics module is designed for use with \SI{160}{\milli\meter} tube length objective lenses, however the actual length of the optics module is less than this, due to space constraints. A `tube length correction' lens is push-fitted into the optics module, at a position such that it forms an image on the sensor at a shorter distance. The magnification was lowered such that the field of view of a standard field number eyepiece (17mm) would match the smaller imaging sensor of the Raspberry Pi Camera v2 (diagonal 4.6mm). An achromatic doublet lens (focal length: 50\,mm, outer diameter: 12.7\,mm, e.g. Thorlabs AC127-050-A) is used to minimize the chromatic aberrations which can occur when using white illumination. By using push fits and screw threads for the lenses, alignment is made as simple as possible, further reducing chromatic and other optical aberrations.

The camera (described further in Section~\ref{sec:camera}) mounts with screws to the bottom of the optics module, with the exposed circuit board protected by a cover.

\subsubsection{Illumination module}

For transmission illumination (Figure~\ref{fig:transmission_illumination_assembly}(b)) the single-print condenser module focuses a white LED onto the sample using a $f= \SI{5}{\milli \meter}$ PMMA condenser lens. It is positioned on a mount which uses a friction dovetail to hold the module in position.  The mount can be positioned in $x$ and $y$ to align the illumination before being secured into position on the main body.  The same mount can be used for different types of illumination module, for example a LED array illumination module can be attached to a frame that connects to the mount.

\subsubsection{Light path}
The OFDS uses a modified version of the OpenFlexure Microscope's brightfield imaging setup~\cite{Collins2020}, adjusting the relative rotations of the camera to the stage. Figure~\ref{fig:all_ray_diagrams}(a) shows the cross section of the brightfield setup.  Illumination from the white LED is collected by the condenser lens and projected onto the sample.  The objective lens and `tube length correction' lens image the sample onto the imaging sensor.

\subsubsection{Sample images -- brightfield}
We have tested the microscope for use with automating blood smear imaging, with a particular focus on blood-borne malaria.  Images of malaria infected human red blood cells, smeared and fixed, were captured using the brightfield setup of the OFDS with a $100\times$, 1.25NA RMS objective lens.

\subsection{Sample preparation and brightfield imaging}
Smears of malaria infected cells were prepared following the standard procedure for lab cultures. Briefly, a culture of \textit{Plasmodium falciparum} malaria parasites (3D7, \SI{3}{\percent} HCT, \SI{5}{\percent} parasitemia, mixed stages) were maintained in human red blood cells and RPMI culture medium, following~\cite{Trager1976,Esposito2010}. The blood was from healthy donors, purchased from NHS Blood and Transplant. They were cultured by adding gas (1\% O$_2$, 3\%CO$_2$) and storing in an incubator at \SI{37}{\degree}.  In ideal conditions, the parasitemia would multiply by 5 every 48 hours. A \SI{200}{\micro \liter} sample of the culture was removed and centrifuged for 20 seconds.  \SI{150}{\micro \liter} of the supernatant was removed and the blood pellet was re-suspended. \SI{5}{\micro \liter} was dropped onto a glass microscope slide and smeared into a thin smear using a second glass slide, following standard thin film smear preparation\cite{HealthOrganization2010}. The smear was dried by hairdryer, and fixed by covering the slide completely with methanol for \SI{30}{\second}.  The methanol was removed and the slide dried again.

\SI{10}{\percent} Giemsa (VWR Giemsa's stain improved solution R66, Gurr\texttrademark, diluted in Evian bottled water, filtered through \SI{20}{\micro \meter} filter) was poured onto the slide and left from \SI{10}{\minute} at room temperature.  The Giemsa was then removed by reverse osmosis under a stream of tap water.  Excess water was removed and the slide dried.

\begin{figure}[t!]
    \centering
    \includegraphics[width=0.8\linewidth]{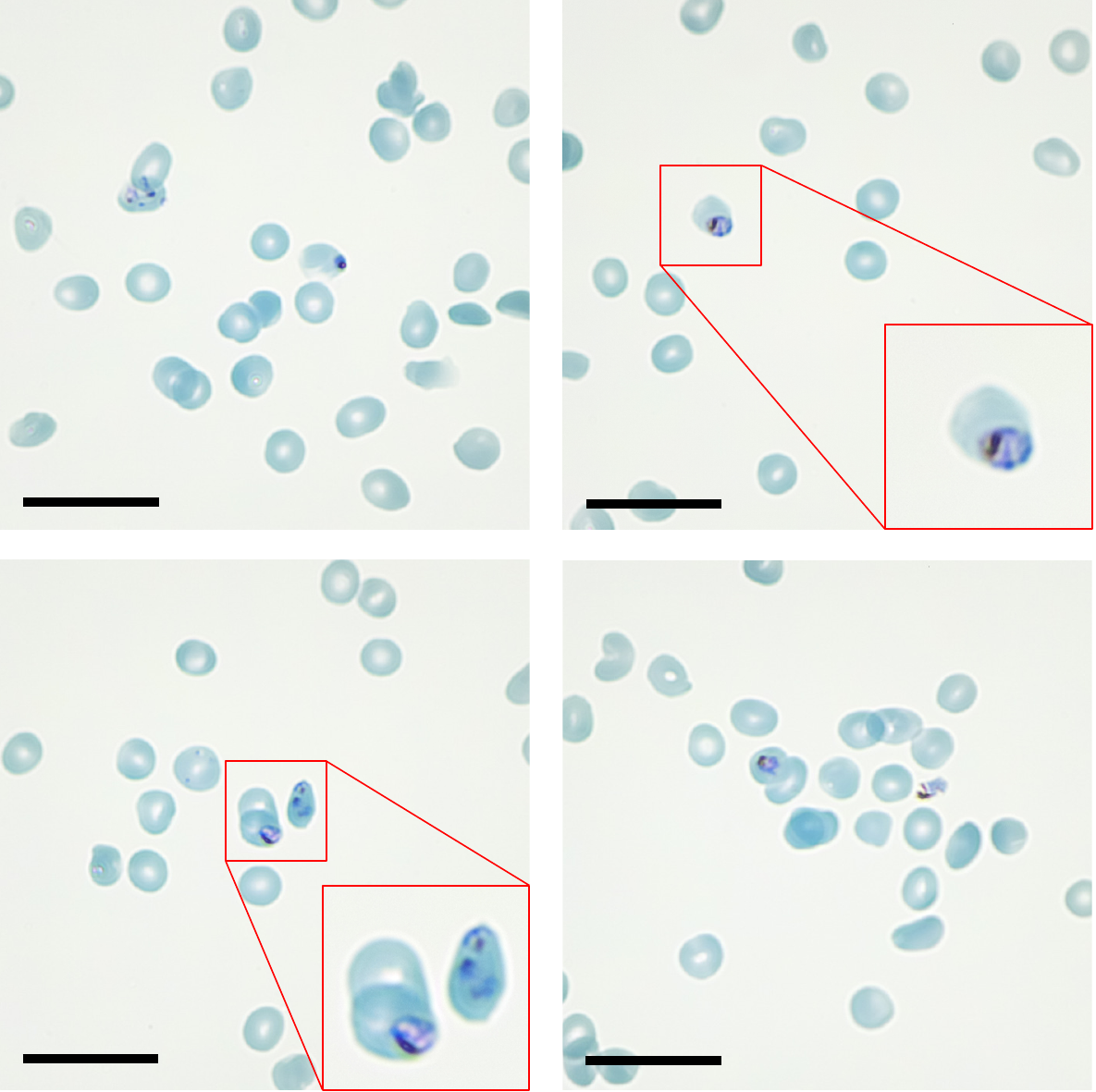}
    \caption{A typical use case of the OpenFlexure Delta Stage is for brightfield imaging of malaria infected red blood cells. These micrographs taken using the OpenFlexure Delta Stage show malaria infected red blood cells, with insets showing zoomed-in regions with the malaria parasites. They were stained using Giemsa, which makes the red blood cells appear blue and the malaria parasites appear purple.  The Raspberry Pi v2 camera (Section~\ref{sec:camera}) captures color images, which is necessary for stained or dyed histological samples. These images were taken using a $100\times$, 1.25NA RMS oil objective lens. Images cropped for visibility, scale bars \SI{20}{\micro \meter}.}
    \label{fig:brightfield-results}
\end{figure}

The slide was imaged using the brightfield configuration of the OFDS with a $100\times$, 1.25NA RMS objective lens. The camera was configured to have an even response across the field of view with the correct exposure time and white balance settings using the OpenFlexure server software\cite{Bowman2020}.  A small drop of immersion oil was placed on the objective lens and the sample was placed on the stage with the blood smear side facing the objective. The sample was moved into a good field of view and focused using the auto focus routine.  Images from the sample can be seen in Figure~\ref{fig:brightfield-results}.

\begin{figure}[t!]
    \centering
    \includegraphics[width=0.8\linewidth]{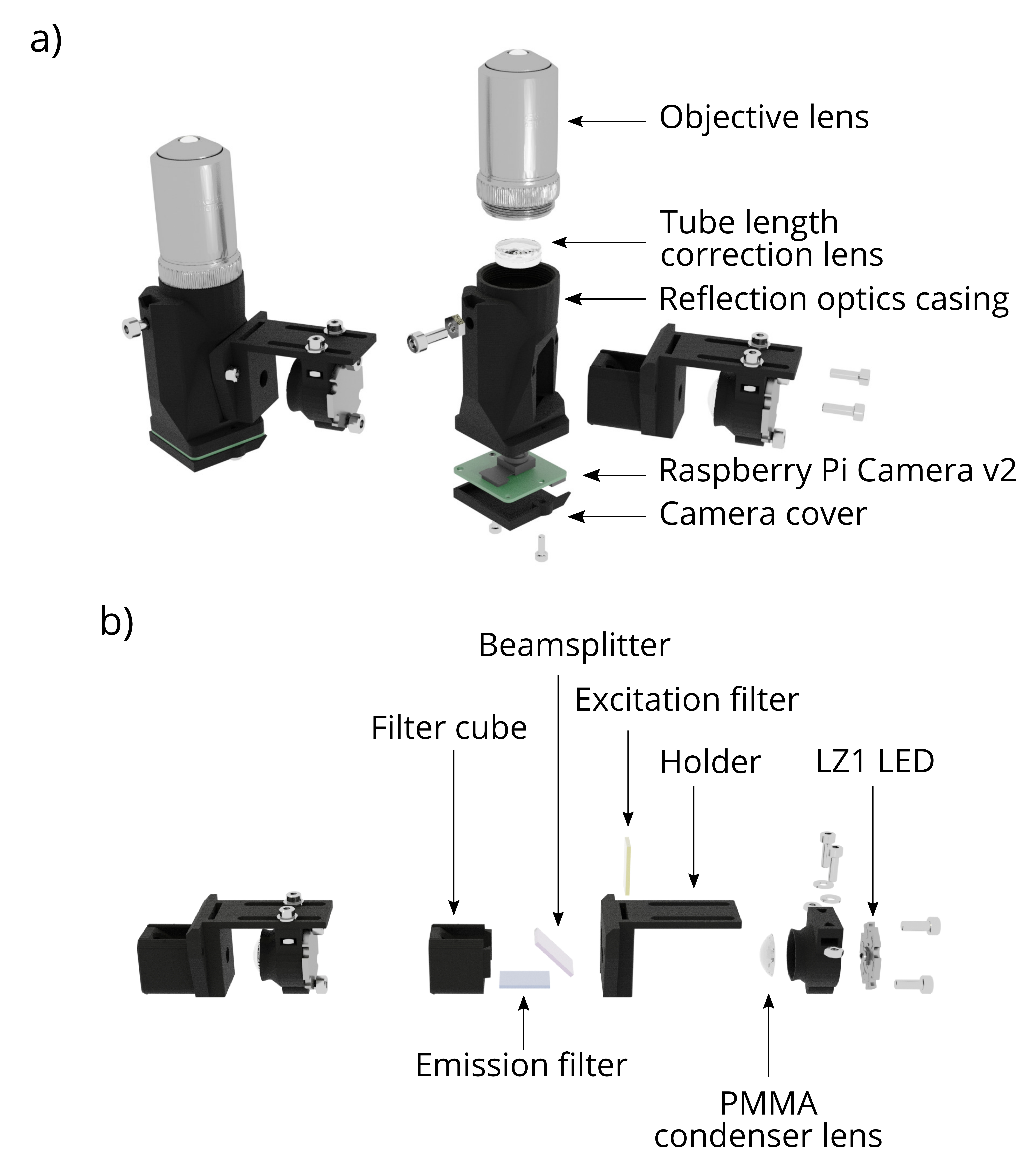}
    \caption{The reflection mode modules uses a modified version of the transmission optics modules. They are constructed from 3D printed parts, custom-cut optical filters, a LZ1 LED and common optical parts. This computer generated render shows the assembly of the reflection (a) optics and (b) illumination modules.}
    \label{fig:reflection-illumination-assembly}
\end{figure}

\subsection{Epi-fluorescence (reflection) illumination}
\label{sec:epi-fluorescence}
\subsubsection{Optics module}
The design of the reflection illumination optics module is similar to the transmission illumination variant.  However, between the `tube length correction' lens and the camera, a square hole is cut out of the side, and there are two screw holes either side (Figure~\ref{fig:reflection-illumination-assembly} (a)). The filter cube attached to the reflection illumination module fits  inside this hole, and it is secured in place with the two screws.

\subsubsection{Illumination module}
The illumination module consists of a filter cube, a condenser module and the reflection illumination holder, (Figure~\ref{fig:reflection-illumination-assembly}(b)).  The filter cube has two slots for custom cut glass filters.  At the bottom of the filter cube is a slot for the emission filter.  At \SI{45}{\degree} to the base there is a slot for the beamsplitter.  The filter cube couples onto the reflection illumination holder with a dovetail, and the combined unit slides into the reflection illumination optics module and is attached with two screws.  In the top of the reflection illumination holder is a slot for a custom cut glass excitation filter. A filter cube, as opposed to a cheaper fluorescent dark field setup, was chosen as it is more optically efficient, and does not photo-bleach the entire sample.

The LED (LED Engin LZ1 with starboard MCPCB) is screwed onto the back of the reflection illumination condenser module. This more powerful mono-chromatic LED was chosen as it is brighter, to increase the signal from the fluorophores.  Inside the reflection illumination module is a pinhole.  A 5\,mm focal length PMMA condenser lens focuses the light from the LED onto the back focal plane of the objective lens.  The condenser module is attached to a slip plate on the reflection illumination holder so that it can be moved to the correct position.

\subsubsection{Light path}
The reflection mode module is used for reflection and epi-fluorescence imaging.  Figure~\ref{fig:all_ray_diagrams} (b) shows the cross section of the reflection imaging set up. The more powerful mono-chromatic LED is attached to the illumination module.  A pinhole is in front of it and the condenser lens projects the illumination from the LED, through the filter cube and onto the back focal plane of the objective lens.  The excitation filter ensures that the illumination wavelength matches the excitation wavelength of the sample's fluorophores. The dichroic beamsplitter reflects the illumination towards the objective.  The objective lens focuses the illumination onto the sample.  The fluorophores are excited and emit illumination of a higher wavelength in all directions.  The objective lens captures some of that illumination and projects it onto the camera sensor.

\subsubsection{Sample images -- epi-fluorescence}
Two samples were used to test two different epi-fluorescence wavelengths. Fluorescent stained malaria infected human red blood cells were imaged using epi-fluorescence.  SYBR Green I (ex = \SI{497}{\nano \meter}, em = \SI{520}{\nano \meter}) was used as the staining agent. This stain binds to double-stranded DNA, such as those present in the malaria parasite. A StarLight calibration slide (Bangs Laboratory SL1GB), consisting of $\approx$\SI{6}{\micro \meter} fluorescent microspheres (ex = \SI{360}{\nano \meter}, em = \SI{450}{\nano \meter}) was also used.

\begin{figure}[!t]
    \centering
    \includegraphics[width=0.65\linewidth]{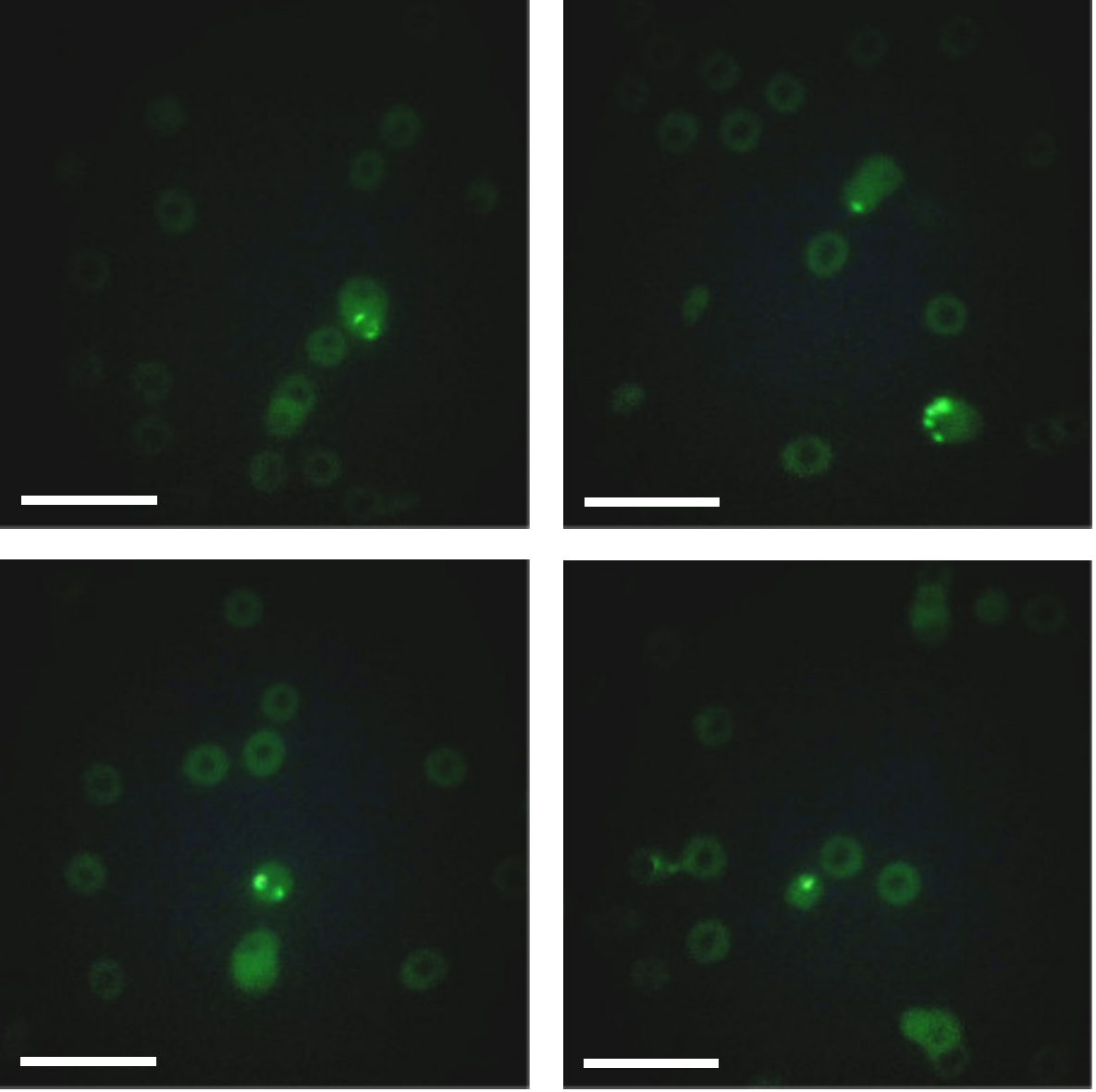}
    \caption{The OpenFlexure Delta Stage is capable of imaging samples with quite low fluorescent intensity. (a) Epi-fluorescence micrographs of SYBR-green stained malaria infected red blood cells show the location of the red blood cells and the brighter regions show the locations of the malaria parasites. It is possible to see the characteristic ring shape of the trophozoite stage of the malaria erythrocytic cycle. (b) Epi-fluorescence micrographs show \SI{6}{\micro \meter} fluorescent dyed $\approx$\SI{6}{\micro \meter}microspheres. They are illuminated with the UV LED and emit blue light.  Images cropped and contrast enhanced for visibility, scale bars \SI{20}{\micro \meter}.}
    \label{fig:fluorescence-results}
\end{figure}

\subsection{Sample preparation and epi-fluorescence imaging}
The sample of malaria parasites in human red blood cells was smeared and fixed as above.  The slide was covered completely with 10\,mM Tris pH8 for \SI{10}{\second}. The excess TRIS was removed, but the slide was not dried.  The slide was then covered completely with 1:3000 dilution of SYBR Green I (Thermofisher, supplied as $10,000\times$ concentrate in DMSO, Ex: 498\,nm, Em: 520\,nm) in 10\,mM Tris pH8.  The slide was enclosed in a dark box for 10\,minutes. The excess SYBR Green 1 solution was removed by reverse osmosis under a stream of tap water. The slide was dried using a hairdryer.

The malaria sample slide was imaged using the epi-fluorescence configuration of the OFDS with a $100\times$, 1.25NA RMS objective lens. A blue LED (LZ1-10B202-0000, $\lambda=460$\,nm) was used, with a short pass excitation filter with 510\,nm transition (Comar optics 510 IK 50), long pass dichroic beamsplitter with 506\,nm transition at \SI{45}{\degree} (Comar optics 550 IY 16), and long pass emission filter with 510\,nm transition (Comar optics 510 IY 50). The images can be seen in Figure~\ref{fig:fluorescence-results} (a).

The microsphere sample slide was also imaged using the epi-fluorescence configuration of the OFDS with a $100\times$, 1.25NA RMS objective lens. A UV LED (LZ1-10UV0R-0000, $\lambda = 365-370$\,nm) was used as illumination. A band pass excitation filter with peak 360\,nm (Comar optics 360 GB 50), a long pass dichroic beamsplitter with 410\,nm transition at \SI{45}{\degree} (Comar optics 454 IY 125) and a long pass emission filter with transition at 435\,nm (Comar optics 435 GY 50) were used in the filter cube. The images can be seen in Figure~\ref{fig:fluorescence-results} (b).

\subsection{Computational illumination with LED array}
\begin{figure}[t!]
    \centering
    \includegraphics[width=0.7\linewidth]{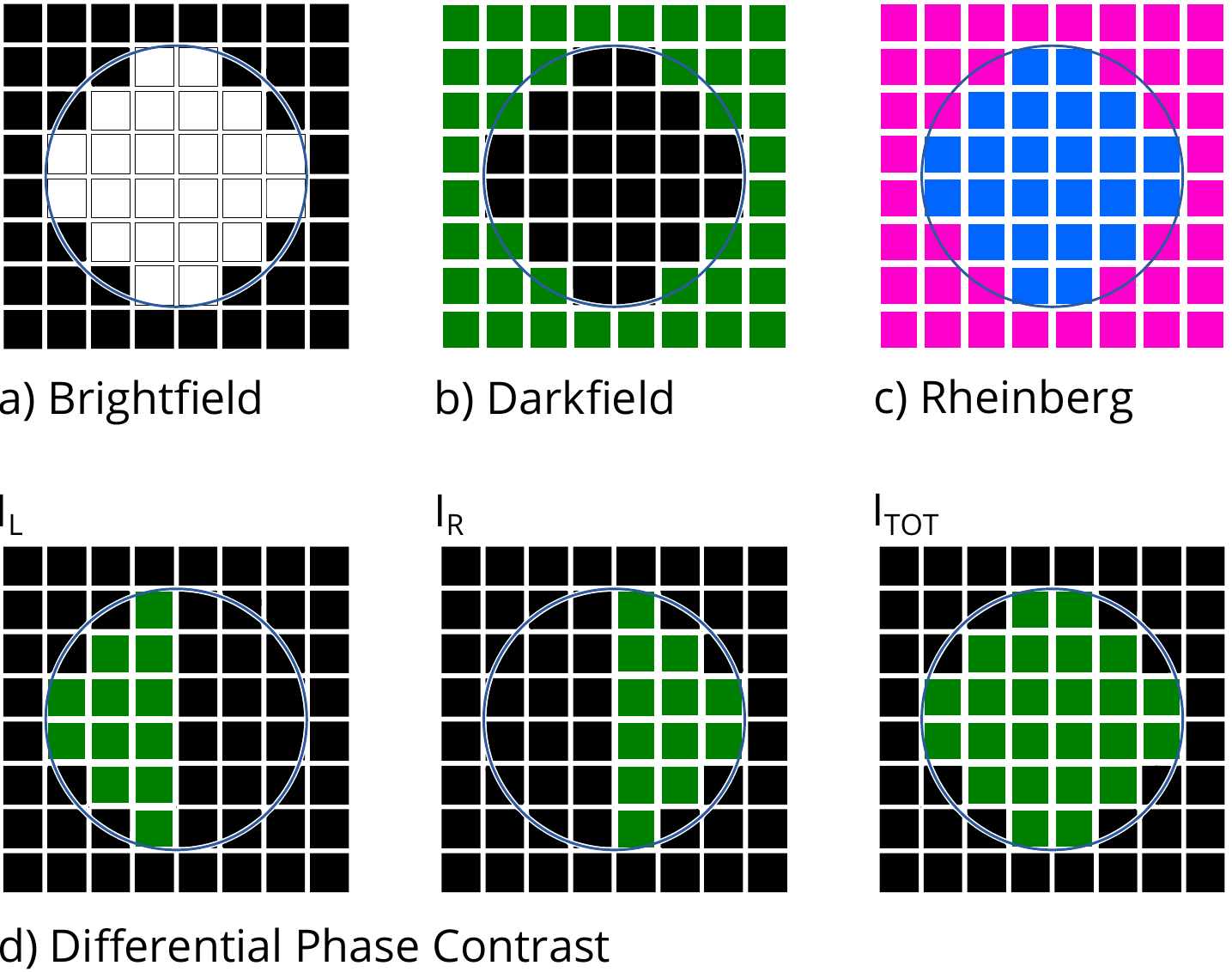}
    \caption{An 8-by-8 LED array (Section~\ref{sec:led-array-info}) is used to provide complex illumination patterns, which can enhance contrast and generate phase images.  This diagram shows examples of LED illumination patterns for different types of imaging.  Each square in the grid represents one LED, and the circle represents an example Numerical Aperture of the objective lens calculated geometrically for a $40\times$, 0.65NA objective , as shown in Figure~\ref{fig:all_ray_diagrams}. The LED array is positioned as close as reasonably possible to the sample ($\approx 5$mm) in order that the NA of the objective falls within the LED array. Each LED `pixel' is 2\,mm $\times$ 2\,mm with a viewing angle of \SI{160}{\degree} and consists of 3 LEDs (red, green and blue) which can be individually controlled to produce a single color. (a) For brightfield imaging, only the pixels within the NA are illuminated, but by illuminating less than the full set, the illumination NA can be reduced. (b) For Darkfield imaging, the LEDS outside the NA are illuminated. (c) For Rheinberg imaging, the `Darkfield' LEDs are illuminated in one color and the `Brightfield' in another. (d) For Differential Phase Contrast imaging, three LED patterns are used, and an image captured for each.  This is automated through the phase contrast extension~\cite{2021LEDRepository}.}
    \label{fig:led_patterns}
\end{figure}

Using a LED array for illumination has been shown to be a low cost way to add new imaging modes to a standard microscope setup~\cite{Zheng2011, Tian2014}. Figure~\ref{fig:all_ray_diagrams}(c) shows the cross-section of the LED array setup for the OFDS.  The LED pixels that fall within the numerical aperture of the objective are shown in blue. The specification and control of the LED array is given in Section~\ref{sec:led-array-info}.  The LED array is positioned such that the imaging NA is one LED pixel smaller than the LED array.  This imaging NA is shown as a  circle in Figure~\ref{fig:led_patterns}(a).

The LED array allows for a spatial illumination source. Using an extension that adds functions to the control software (Section \ref{sec:extensions})~\cite{2021LEDRepository} it is possible to control each of the 64 RGB LEDs individually.  As the LED array is a flat surface, the LEDs closest to the objective lens (in the centre of the array) will appear brighter than those further away (at the corners).  To correct for this, the LED uses a look up table (LUT) based on the height of the LED array and the LED pixel density to make an illumination profile that simulates a curved surface.  This means that the intensity of the illumination of any pixel at the pupil of the objective lens will be equal.

\begin{figure}[t!]
    \centering
    \includegraphics[width=0.6\linewidth]{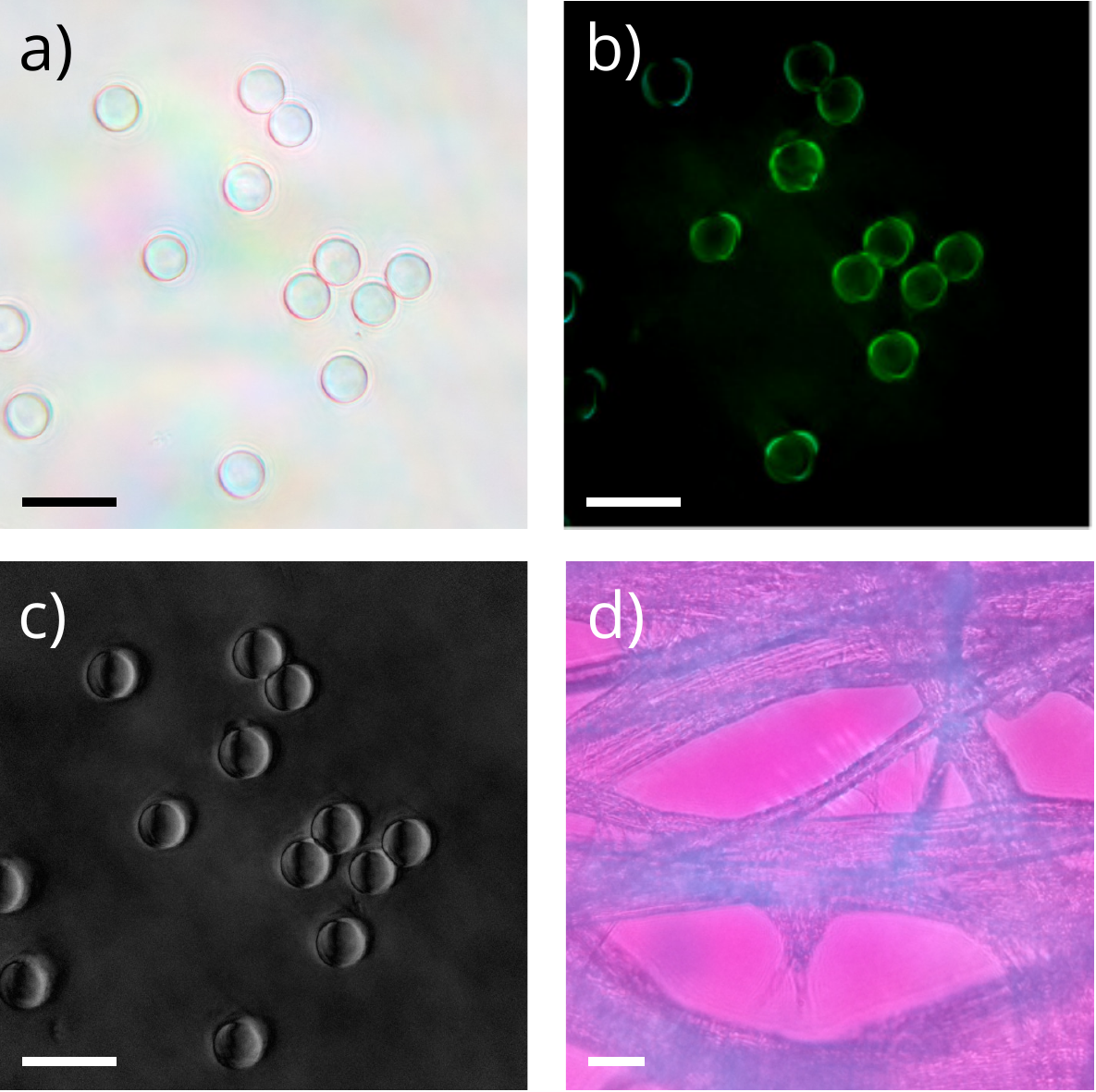}
    \caption{Sample images taken with the different imaging modes using the LED array. (a) Brightfield image of \SI{9.56}{\micro \meter} SiO$_2$ microparticles (RI: 1.43) suspended in Milli-Q water (RI: 1.33) using the central $2\times2$ LEDs. ($63\times$, 0.90NA) (b) Darkfield image of \SI{9.56}{\micro \meter} SiO$_2$ microparticles.  ($63\times$, 0.90NA, Imaged using green LEDs, Contrast enhanced). (c) Differential Phase Contrast image of  \SI{9.56}{\micro \meter} SiO$_2$ microparticles ($63\times$, 0.90NA) showing the phase gradient along the horizontal direction.
        (d) Rheinberg image of tissue paper ($40\times$, 0.65NA). All scale bars \SI{20}{\micro \meter}.}
    \label{fig:led_sample_images}
\end{figure}

\label{sec:LEDbrightfield}
For brightfield imaging with the LED array, all the pixels within imaging NA circle are lit as shown in Figure~\ref{fig:led_patterns} showing that the illumination NA is equal to the imaging NA.  The illumination NA can be reduced by reducing the radius of the illuminated pixels.  As the illumination NA is reduced, the coherence is increased and the depth sectioning is reduced, as shown in Figure~\ref{fig:led_sample_images}(a).

\label{sec:darkfield}
In darkfield microscopy~\cite{Zheng2011}, the unscattered illumination is excluded from the sensor. If the LEDs outside of the imaging NA are illuminated, then there will be no direct light entering the objective, and the image will be dark.  When a sample is loaded in position, it scatters the light from the LEDs into the objective and forms a darkfield image as shown in Figure~\ref{fig:led_sample_images}(b). The darkfield LEDs were chosen to be only green as they have the highest brightness (while minimising chromatic aberrations) and the Bayer filter pattern of the camera sensor is more sensitive to green.

\label{sec:rheinberg}
Rheinberg imaging~\cite{Rheinberg1896} follows the same principle as darkfield imaging, but because it is possible to change the colors of the LEDs, the LEDs that would form the brightfield image (within the imaging NA) can be one color, and the LEDs that would form the darkfield image (outside the imaging NA) can be a different color, as shown in Figure~\ref{fig:led_patterns}(c). The images produced by Rheinberg imaging are a form of optical staining.  The background of the image will be in the brightfield color, and the darkfield color will provide enhanced contrast effects.  An example image using this technique is shown in Figure~\ref{fig:led_sample_images}.

\label{sec:dpc}
Differential Phase Contrast (DPC) can be achieved using the brightfield LEDs of the LED array~\cite{Tian2014}.  The principle of DPC is that the gradient of phase can be calculated from two intensity images when the sample is illuminated from opposite angles.  Figure~\ref{fig:led_patterns}(d) show the LED patterns when illuminated from the left ($I_\text{L}$) and right ($I_\text{R}$).  $I_\text{TOT}$ is the brightfield image ($I_\text{L} + I_\text{R}$). The DPC image is calculated as
\begin{equation}
    I_\text{DPC}=\frac{I_\text{L}-I_\text{R}}{I_\text{TOT}}.
\end{equation}

This DPC image is related to the object's phase gradient along the axis of asymmetry\cite{Hamilton1984}, creating a `shadow' effect along the axis. DPC images are automated on the OFDS using the phase contrast extension~\cite{2021LEDRepository}.  Figure~\ref{fig:led_sample_images} shows a sample image captured using the DPC technique.

With the use of extensions, as described in Section~\ref{sec:extensions}, many more illumination routines such as 3D differential phase contrast~\cite{Tian2014}, color-multiplexed differential phase contrast~\cite{Phillips2017}, or Fourier ptychography~\cite{Zheng2013} are possible.

\section{Stage and body design}
The OFDS is majority 3D printed.  The main body, which consists of a 3D moving stage, is printed as one piece. The base of the stage houses the electronic components.

\subsection{Stage mechanism}
One of the most important aspects of a microscope's design is a stable stage. A high resolution microscope with a stage that drifts or shakes will produce low quality images. The microscope's stage is inspired by a delta robot design~\cite{Rocholl2012}. The sample stage is mechanically controlled, and can move the sample in three dimensions. This design is particularly advantageous as the optics remain stationary.

\begin{figure}[t!]
    \centering
    \includegraphics[width=0.8\linewidth]{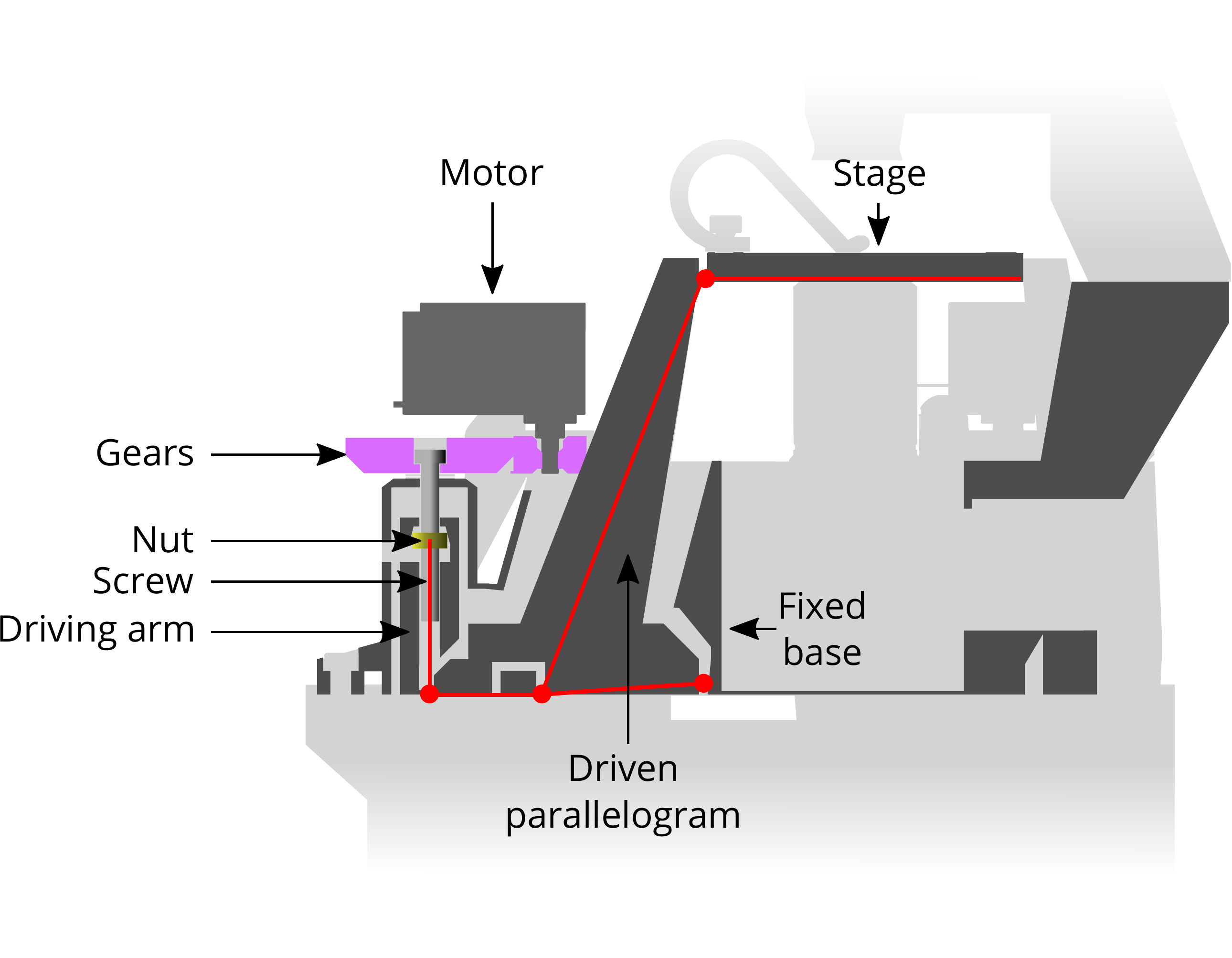}
    \caption{The stage is 3D-printed as a single part, and uses the flexure properties of the plastic to enable fine positioning of the stage. This cross section shows this mechanism used to control the actuators and move the stage with its pin-joint structure shown in red.}
    \label{fig:actuator}
\end{figure}

The kinematics of the OFDS are based around the delta robot design. The delta robot is a type of parallel arm robot. As shown as a cross-section in Figure~\ref{fig:actuator}, it consists of a fixed base and a moving stage. The moving platform is situated directly above the fixed platform.  The kinematic chain consists of a driving arm and a driven arm. There are three of these, situated at \SI{120}{\degree} around the platforms.  The driven arm is a parallelogram, taking advantage of the well characterised flexure mechanism of the 3D printed plastic~\cite{Sharkey2016}. As per that tested design, the thickness of the flexure is set to three layers of plastic, with a flexure length of 1.5mm.  A compromise of deflection range and stiffness, this design allows approximately \SI{6}{\degree} of motion without exceeding the elastic limit. By using parallelograms, the movement of the stage is restricted to solely translation.  This means that although there may be some parasitic z-motion for x and y movements, it does not tilt or rotate, undesirable attributes for a microscope stage.

The three driving arms are linear actuators, controlled by three stepper motors. The actuator design, shown in Figure~\ref{fig:actuator}, is adapted from the OpenFlexure Microscope~\cite{Collins2020, Sharkey2016}. The stepper motor drives the 3D printed actuator mechanism in a vertical motion. As the actuator is connected to the parallelogram structures by flexures, they are driven by the vertical movement of the actuator.

\begin{figure*}[t!]
    \centering
    \includegraphics[width=\linewidth]{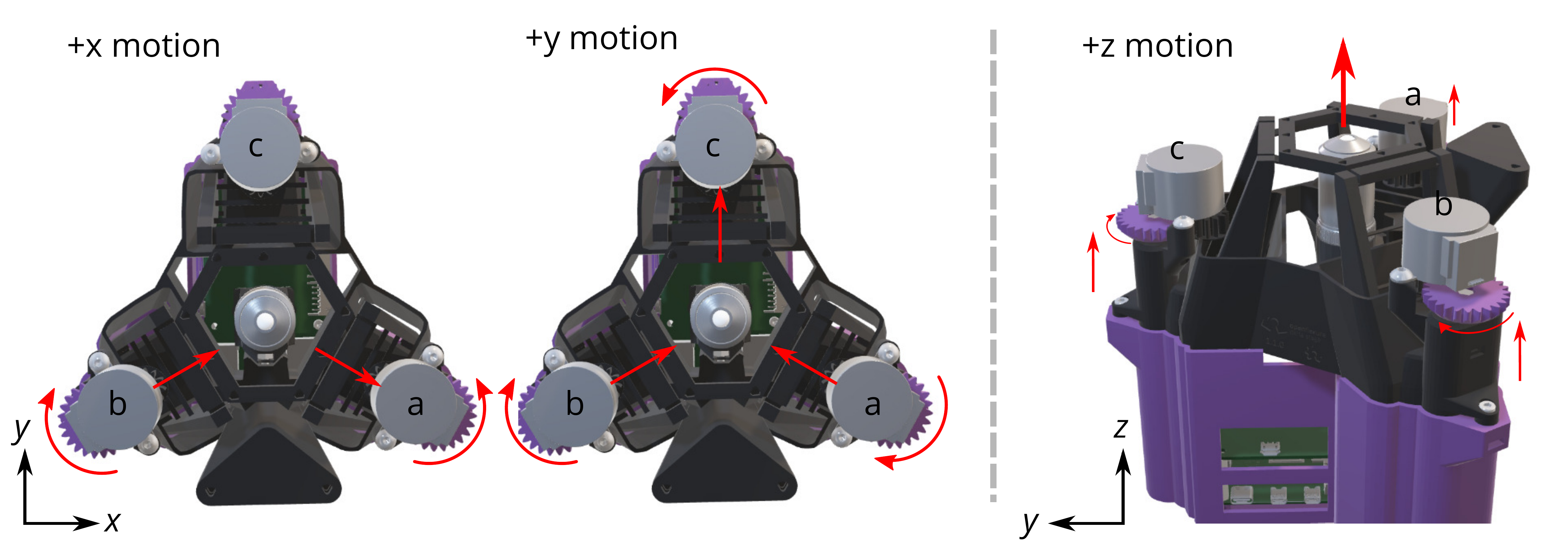}
    \caption{The Cartesian stage motions are created from the rotations of a combination of the actuator gears. For example, to move the stage in the positive $x$-direction, gear `b' rotates clockwise and gear `a' rotates anti-clockwise.}
    \label{fig:delta_robot_kinematics}
\end{figure*}

There has been some work done in calculating the kinematics of delta robot mechanics~\cite{Li2019}.  However, as the driven arms are constrained to only bend through small angles (\SI{6}{\degree}), the small angle approximation means that the kinematics can be approximated using linear relations. It would be possible for users to modify the driven arms to increase the range of motion, but this will come at a reduction in stability and the small angle approximation will no longer be applied. As shown in Figure~\ref{fig:delta_robot_kinematics}, the movement of each of the parallelograms can be mapped from their delta-bot kinematics to Cartesian kinematics, which is more intuitive for controlling the stage.  The software converts the Cartesian kinematics into the delta kinematics using the following relation (in addition to $x,y,z$ scaling):
\begin{equation}
    \begin{bmatrix}
        x \\
        y \\
        z
    \end{bmatrix}
    = \begin{bmatrix}
        -\cos{30} & \cos{30} & 0  \\
        \cos{60}  & \cos{60} & -1 \\
        1         & 1        & 1
    \end{bmatrix}
    \begin{bmatrix}
        a \\
        b \\
        c \\
    \end{bmatrix}
\end{equation}

The geometric proportions of the stage mechanism have been optimised to allow for a stable stage with the necessary movement accuracy, travel, and size. The horizontal travel is set by the stage height, as shown in Figure~\ref{fig:actuator}. This is a compromise between the overall height of the structure--affecting stiffness and print reliability and travel--which increases with a taller stage. The vertical travel is set by the lever length--the distance between the central pivot and the point where the legs attach to the actuator levers. The length of the actuating lever sets the number of motor stpdf across the range of travel. Starting with the actuators in their central positions, the range of the stage's motion is [9mm, 12mm, 5mm].  Further information about the parameters in the design of the stage can be found in the online documentation~\cite{2022OpenFlexureDocumentation}.

It is possible to mount different samples onto the stage using stage adapters which screw onto the top of the stage.  Sample clips are used to hold microscope slides with fixed samples, whereas \SI{35}{\milli \meter} and \SI{55}{\milli \meter} Petri dish adapters enables the imaging of live samples.

\subsection{Base}

The base is at the bottom of the microscope.  It is designed to keep the electronic components securely in place and provide a stable platform for the main body of the microscope.  The Raspberry Pi sits in the bottom of the base, on four raised 3D printed standoffs.  There are holes in the side of the base to allow connections and there is a cutout underneath to access the SD card.  \SI{20}{\milli \meter} M2.5 standoffs then secure the Raspberry Pi to the base.  The Sangaboard then sits on top of these standoffs and is secured with M2.5 screws.  Again, there are holes for access to the Sangaboard connectors.

The main body of the microscope fits tightly inside the base and its feet sit on elevated platforms. It is secured to the base by three screws connected to each actuator column.  It is recommended that rubber feet are attached to the bottom of the base, or that it sits on a layer of foam to reduce environmental vibrations.

\section{Electronics}
The OFDS forms the basis of a digital `smart' microscope, capable of automating the imaging tasks. To operate,  it requires some basic electronics and electrical components.

\subsection{Electronic components}
\subsubsection{Illumination}
\paragraph{Transmission LED}
A white through hole 5\,mm 3.2V LED (Nichia NSPL500DS) is used for transmission illumination.  This LED was chosen because it has a spectrum that closely matches that of a filament bulb, but any 5\,mm LED can be used. A \SI{100}{\ohm} resistor is soldered to one of the legs of the LED and it is connected to the 5V and ground connectors of the Raspberry Pi's GPIO.  This means that whenever the Raspberry Pi is turned on, the illumination will turn on.

\paragraph{Epi-fluorescence LED}
LED Engin LZ1 LEDs with the star MCPCB are used for illumination. This was chosen as it provides the high flux required for fluorescence illumination in a convenient form factor and with wide availability.  They are available in 5 colors (Table~\ref{tab:LED_colours}) which are suitable for a range of fluorophores and for epi-illumination. They are attached to an aluminium backplate for heat dissipation. They require a bench top constant-current power supply, although it may be possible in the future to specify a reliable LED driver.

\begin{table*}[t!]
    \caption{The LZ1 LEDs were chosen for their brightness and their range of colors. Their wavelengths are applicable for a wide variety of reflection illumination samples.}
    \centering
    \begin{tabular}{@{}lrl@{}}
        \toprule
        \textbf{Color} & \textbf{Wavelength (nm)} & \textbf{Manufacturer number} \\ \midrule
        UV             & 365-370                  & LZ1-10UV0R-0000              \\
        Blue           & 460                      & LZ1-10B202-0000              \\
        Green          & 523                      & LZ1-10G102-0000              \\
        Red            & 623                      & LZ1-10R102-0000              \\
        Cool white     & 5500K                    & LZ1-10CW02-0055              \\ \bottomrule
    \end{tabular}
    \label{tab:LED_colours}
\end{table*}

\paragraph{LED array}
\label{sec:led-array-info}
\begin{figure}[t!]
    \centering
    \includegraphics[width=0.8\linewidth]{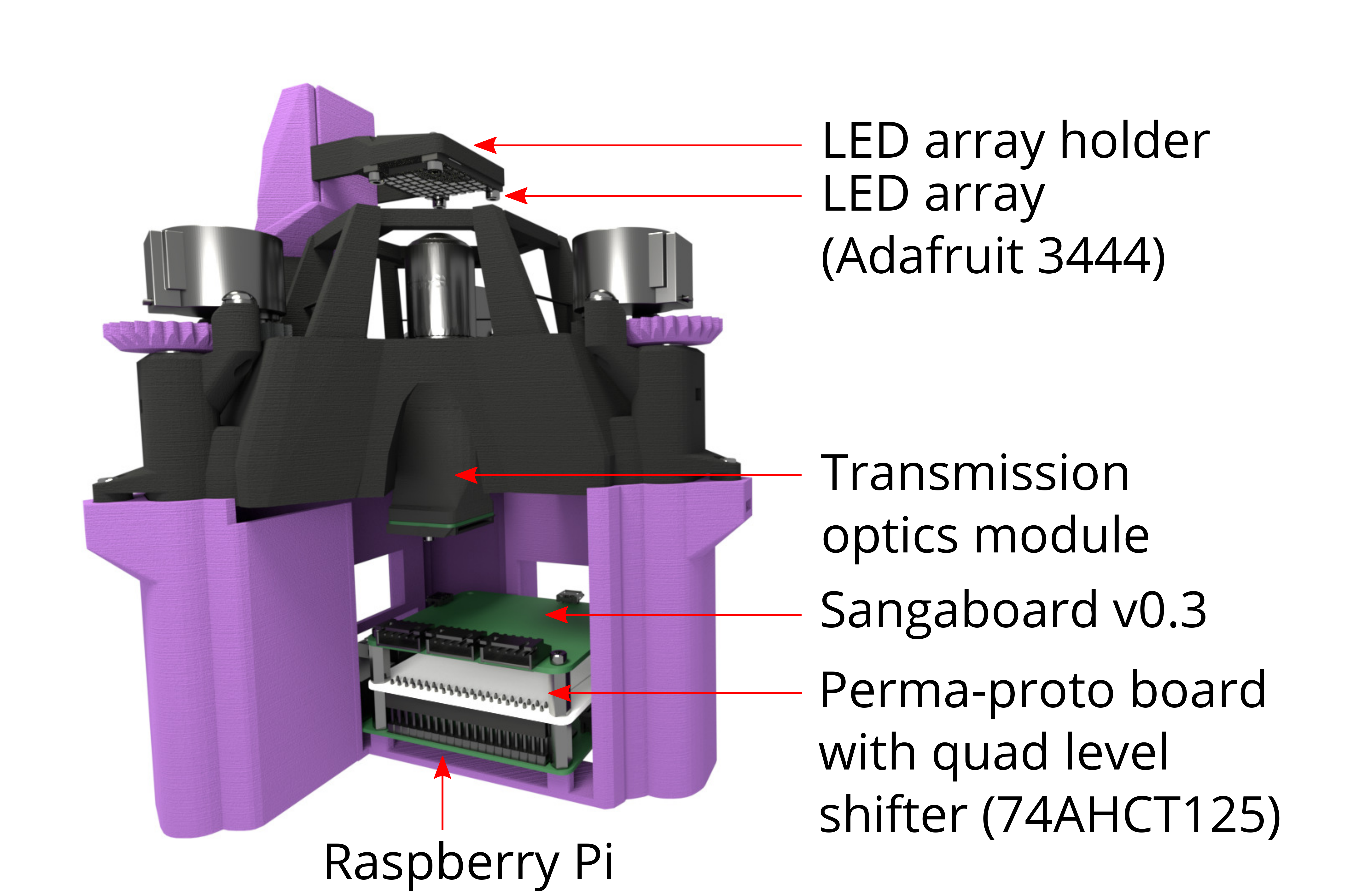}
    \caption{The electronics and illumination holder are modified for the LED array, but it is still possible to use the same main body and base as in the other imaging modalities.  This computer generated render of the OpenFlexure Delta Stage shows, with  the base cutaway,  how the extra components are assembled.}
    \label{fig:led_array_setup}
\end{figure}

The LED array (Adafruit 3444) consists of 64 DotStar (APA102-2020) LEDs spaced 3.5\,mm over an eight by eight grid. This commercially available LED array has a small enough pitch suitable for the scale of the OFDS with a convenient control interface.  Each LED `pixel' is 2\,mm $\times$ 2mm with a viewing angle of \SI{160}{\degree} and consists of 3 LEDs (red, green and blue) and can be addressed individually. The array is connected to the Raspberry Pi GPIO pins via a 3V to 5V quad logic level shifter (74AHCT125) mounted on a perma-proto board. It is mounted to the illumination arm using a 3D printed mount. In future, a condenser lens should be added to the mount, as demonstrated in~\cite{Tian2014}, in order that the pixels at the edge of the LED array can have more of their intensity captured by the objective lens. The LEDs are controlled using the Adafruit Circuit Python DotStar driver library~\cite{2021AdafruitRepository}. An interface to the OpenFlexure Microscope Server's interface is provided with an extension~\cite{2021LEDRepository}, which presents a GUI to the user, allowing them to turn on and off individual and groups of LEDs, as well as run different imaging routines.  These routines synchronise the illumination of LEDs with the camera, allowing for complex imaging techniques such as differential phase contrast imaging. The specification of the LEDs is shown in Table~\ref{tab:LED_array_specification}.

\begin{table*}[b!]
    \caption{Specification of the RGB LEDs in LED array (Adafruit 3444).}
    \label{tab:LED_array_specification}
    \centering
    \begin{tabular}{@{}lrr@{}}
        \toprule
        \textbf{Color} & \textbf{Wavelength (nm)} & \textbf{Brightness at 20mA (mcd)} \\ \midrule
        Red            & 620-625                  & 300-330                           \\
        Green          & 520-525                  & 420-460                           \\
        Blue           & 465-470                  & 160-180                           \\ \bottomrule
    \end{tabular}

\end{table*}

\subsubsection{Camera}
\label{sec:camera}
The camera at the bottom of the optics module is a Raspberry Pi Camera Module v2~\cite{2021Raspberry2}. This uses a well characterised Sony IMX219 sensor, and connects to the Raspberry Pi with a ribbon cable.    It is an 8 megapixel ($3280 \times 2464$), back illuminated CMOS color image sensor.

The optics module creates an image of the sample directly onto the sensor, so the camera's attached lens is not required and it can be unscrewed using a plastic tool. However, as the sensor has been optimised in optical design and firmware for its original short focal lens, it now generates images with a non-uniform background and color response.  This effect has been considered recently, and can be calibrated and corrected on the fly using software~\cite{Bowman2020}.

\subsubsection{Motor controller and motors}
The OFDS has a 3 axis stage.  Due to the complexity of the geometry, it is not possible to be controlled by hand and so requires 3 stepper motors to actuate the stage's levers to enable precise movements.
28BYJ-48 stepper motors were chosen as they are low-cost, low-power and light. They are unipolar and require a convenient 5V power supply.  Due to the reduction gearing, they are slower than typical Nema-17 motors, but have sufficient speed for the microscope stage movement.  Additionally they can suffer from backlash, but this is mitigated in the control software. They are accurate enough that it is possible to move and focus objects using a step size of half a micron when using a $100\times$ objective. They are controlled by a Sangaboard (v0.2 or v0.3)~\cite{2021SangaboardRepository}. This is an open-source, custom 3-axis motor controller board built around an Arduino architecture.  It connects to the Raspberry Pi via USB and has three connections for each motor. It is necessary to power this controller board from a separate 5V power supply so as not to overdraw current from the Raspberry Pi.

\subsection{Computing}
\begin{figure}[!h]
    \centering
    \includegraphics[width=0.8\linewidth]{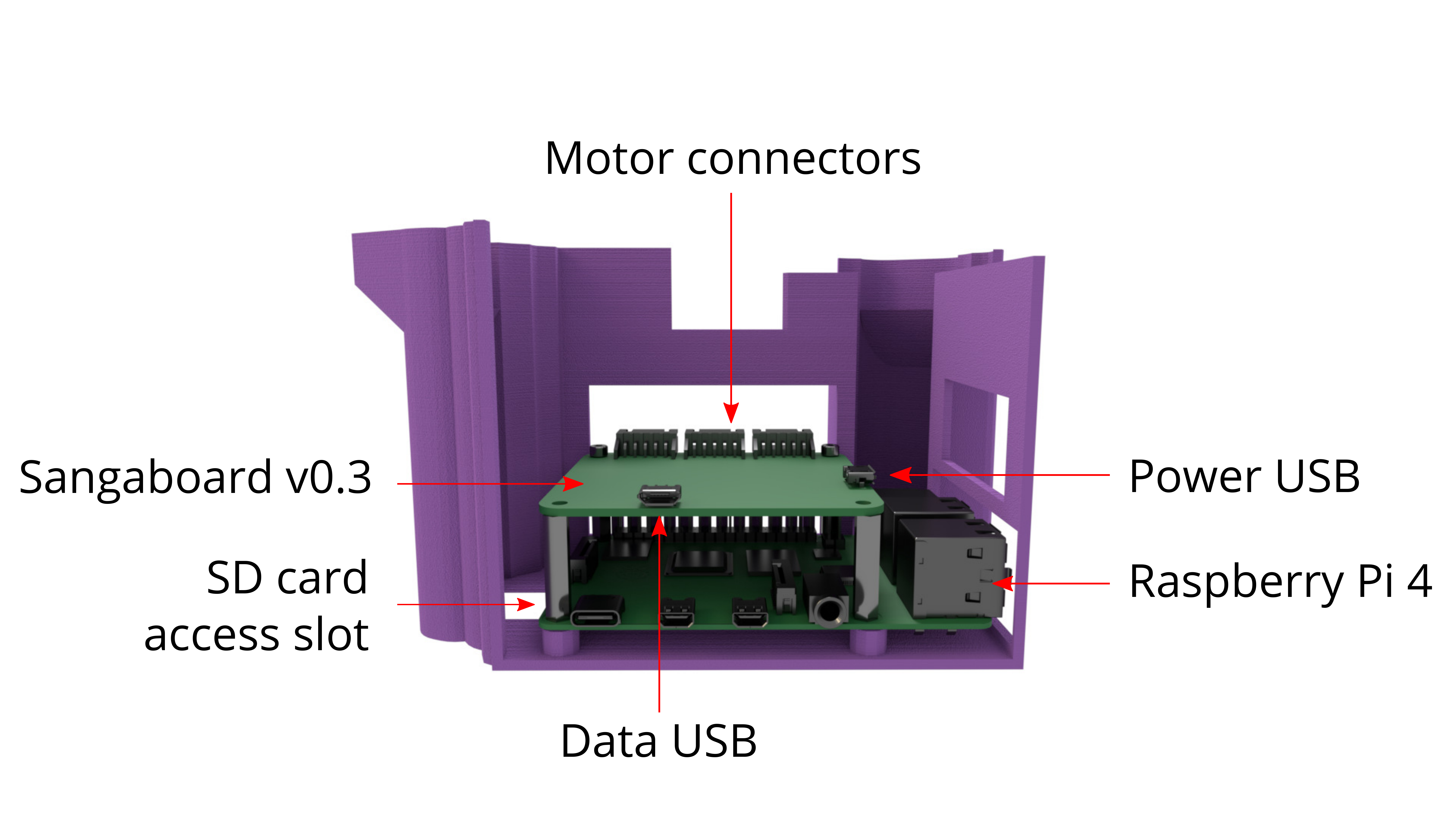}
    \caption{The computing parts are housed tidily in the base of the OpenFlexure Delta Stage. This computer generated render demonstrates a cutaway of the base of the OFDS, showing the positions of the Raspberry Pi and Sangaboard and their access holes.}
    \label{fig:base_cutaway}
\end{figure}

\subsubsection{Raspberry Pi}
In the base of the OFDS is a Raspberry Pi, as shown in Figure~\ref{fig:base_cutaway}.  This controls all the electronic components and can be used as a data store for the images and videos. A Raspberry Pi 4 is recommended over previous models, due to its more powerful processor, large RAM and still modest cost.  The software used is the OpenFlexure Microscope software  (2.10)\cite{collins2021, 2021OpenFlexureRepositoryc}. This python-based software uses a server-client architecture. This means that the OFDS can be controlled either locally on the Raspberry Pi, or remotely over a http connection.  This is distributed as a pre-built Raspberry Pi OS SD card image for ease of use and to provide users with a familiar interactive operating system. The motors are controlled using a Sangaboard v0.3~\cite{2021SangaboardRepository} which is positioned above the Raspberry Pi on 20\,mm standoffs.

The Raspberry Pi Camera v2 is connected directly to the Raspberry Pi using a ribbon cable into the camera port. The Raspberry Pi Camera v2 was chosen because of its excellent integration with the Raspberry Pi. The OpenFlexure Microscope software communicates with the camera using the \texttt{picamerax} library~\cite{2021Picamerax}.  This library takes advantage of the camera's GPU based imaging pipeline to provide additional functionality such as auto gain and exposure, white balance and flat field correction with minimal processing time.  If the OFDS is being used as a stand alone device, a USB keyboard and mouse can be attached and a display can be connected via HDMI.

\subsubsection{Core software}
The core software of the microscope, as described in \cite{collins2021}, is a python based flask app.  The software controls the hardware (camera and motors), performs basic imaging processing (camera calibration), launches a server and presents a web API.  Users can connect to the microscope via this API, to view a live stream from the cameras, control the hardware and capture images.  This means it is possible to control the microscope from on-board the Raspberry Pi, or from an external client computer over the internet. For most users, this is done using from a web browser or OpenFlexure Connect, an electron app. For more advanced users, the on-device scripting extension\cite{2021OpenFlexureRepositoryd}, the OpenFlexure Python Client\cite{2020OpenFlexureRepository} and the OpenFlexure MATLAB client\cite{2021OpenFlexureRepositoryb} allow scripts to send commands directly to the OFDS. This flexibility suits the OFDS when it is placed in Microbiological Safety Cabinets (MSCs) or incubators, so that the researchers can control the device remotely. It is also possible to run complex routines on the microscope, such as autofocusing,  tiling scans, triggering imaging and stage movement events based on image analysis.

\subsubsection{Scripting automation through `Extensions'}
\label{sec:extensions}
It is possible to add ``extensions'' to the OpenFlexure server software.  This allows users to create additional functionality and to control additional hardware.  It is a minimal amount of python code that is executed on the server, as if it were part of the core software.  Extensions can create their own web APIs and GUIs to enable users to activate certain additional functionality.

The software to control the LED array was written as an extension~\cite{2021LEDRepository}.  Using the extension functionality and the Adafruit drivers~\cite{2021AdafruitRepository}, the extension presents a simple interface to run routines such as dark field imaging and phase imaging.

\section{Engineering considerations}
The OFDS was carefully designed to meet our objective that the device is convenient for researchers to replicate, at a reasonable cost, while including features that are useful for general experimentation.

\subsection{Design considerations}
The optical design of the microscope is conventional, prioritising quality and convenience over extreme cost minimisation. The brightfield version described here costs $\approx$£244 (\$336). The device is instantly recognisable as a microscope, providing an intuitive experience for new users. The most expensive parts of the OFDS are the optics and the electronics. By specifying RMS objectives, we enable researchers to reuse old objective lenses, buy them second-hand or purchase them for a low cost. The electronics chosen are also convenient to source, well documented, and affordably priced. The hardware chosen to assemble the device are commonly available machine screws, nuts and standoffs. Although the device can be used for different purposes, it has been designed so that the choices are not overwhelming for the end user.  All imaging modes use the same main body, base, and illumination dovetail. This simplifies the design and makes it easier for users to change modes. The design has been refined over many iterations, as we have learned from end users about their typical use cases and issues, with careful consideration made to quality of life improvements.

\subsection{Manufacturing considerations}
The device is designed to be manufactured by the end user. It was very important therefore to make sure that it can be printed and assembled by someone without a industrial workshop or a large amount of manufacturing experience.  3D printing itself has opened up a new world of distributed manufacturing, and the designs of the 3D printed parts were chosen to be able to be printed on a range of 3D printers. They require no supports, meaning clean-up after printing is easy.  There are also not a large number of parts, making manufacturing easier. For example, the main body is printed as one single piece, making it easier to replicate. The device does not require expensive or complicated tools for assembly, instead it can be put together using hand tools. We have listened to the community about issues when assembling the device, and made improvements to improve the experience.

\subsection{Development infrastructure and continuous deployment}
The 3D models are designed using OpenSCAD~\cite{2021OpenSCAD} and version control is managed using git and GitLab.  Because the OpenSCAD code is parameterised, it is intuitive to adapt and change the dimensions of the designs to suit different requirements. Using the principles outlined in \cite{Stirling2022}, GitLab's continuous integration and continuous development (CI/CD) tools are used to automatically compile the OpenSCAD code into STL files, and store them alongside the assembly instructions for download from the server.

\subsection{Documentation}
Documentation is key to the success and replication of designs. GitBuilding~\cite{2021Gitbuilding} is used to generate detailed and high quality instructions for printed and assembling the OFDS, together with a bill of materials, photos and links to download the STLs~\cite{2022OpenFlexureDocumentation}.

\subsection{Open Source Hardware philosophy}
At the heart of the OpenFlexure philosophy is the open-source nature of our projects.  The code for both the software and hardware has been available online since the project's inception and its design has been shaped because of its open source nature. It now has an active community of users, who use it for their own scientific experiments and instrument designs. Users from around the world have already manufactured and used the OFDS, and have suggested improvements and features they would like to use through the GitLab repository~\cite{2021OpenFlexureRepository}, our forum~\cite{2021OpenFlexureForum} and social media. This iterative feedback process has meant the OFDS has a track record of a reliable and replicable design, and so it should be actively supported into the future. Being open-source has helped it to become useful for more use cases than the maintainers could ever develop and test.

\section{Conclusion}
In this paper, the overall design of the OFDS has been discussed and rationalised.  The construction and operation of the different optics modes have been explained, and some sample images of each of the different modes have been presented.  The mechanics and kinematics of the `delta robot' stage have been described and justified. The efficacy of the open-source nature of the project has been shown, alongside its development pipeline and CI/CD tools. The appeal of the OFDS for biologists and physicists has also been outlined, with some use cases.

Future work will include the development of more extensions to control the LED array, to enable even more imaging applications. The optics module will also be improved to allow easier objective lens and filter cube switching.  The community will continue to be listened to and improvements made, based on real world applications of the device.

\section*{Funding}
This work was funded by the Engineering and Physical Sciences Research Council [EP/R013969/1, EP/R011443/1]; Royal Society [URF\textbackslash R1\textbackslash 180153, RGF\textbackslash EA\textbackslash 181034]

\section*{Data Availability}
All materials and data are available at: \url{https://gitlab.com/openflexure}. All the original data is on a Zenodo archive at:
\url{https://doi.org/10.5281/zenodo.6225993}. The archived versions of the software and hardware described in this paper are available on Zenodo at: \cite{Collins2021b} and \cite{OpenFlexureFiles}.

\section*{Disclosures}
The authors declare no conflicts of interest.

\bibliographystyle{unsrt}
\bibliography{references}

\end{document}